\documentclass[journal]{new-aiaa} 
\usepackage[utf8]{inputenc}

\usepackage{graphicx}
\usepackage{color,bm}
\usepackage{amsmath}
\usepackage[version=4]{mhchem}
\usepackage{siunitx}
\usepackage{longtable,tabularx}
\definecolor{green}{rgb}{0.4660, 0.6740, 0.1880}

\title{
Information-theoretic machine learning for time-varying mode decomposition of {\color{black}separated aerodynamic flows}
}

\author{
Kai Fukami\footnote{Corresponding author, Associate Professor, kfukami1@tohoku.ac.jp, Young Professional Member.}
}
\affil{Department of Aerospace Engineering, Tohoku University, Sendai, 980-8579, Japan}

\author{
Ryo Araki\footnote{Assistant Professor, araki.ryo@rs.tus.ac.jp}
}
\affil{Department of Mechanical and Aerospace Engineering, Tokyo University of Science, Noda, 278-8510, Japan}

\begin{document}

\maketitle

\begin{abstract}
{
We perform an information-theoretic mode decomposition for {\color{black}separated aerodynamic flows}.
The current data-driven approach based on a neural network referred to as deep sigmoidal flow enables the extraction of an informative component from a given flow field snapshot with respect to a target variable at a future time stamp, thereby capturing the causality as a time-varying modal structure.
We consider {\color{black}four} examples of separated flows around {\color{black}a wing}, namely, 1. laminar periodic wake at post-stall angles of attack, strong {\color{black}gust-wing} interactions of 2. numerical and {\color{black}3. experimental measurements}, and 4. a turbulent wake in a spanwise-periodic domain.
The present approach reveals informative vortical structures associated with a time-varying lift response.
For the periodic shedding cases, the informative structures vary in time corresponding to the fluctuation level from their mean values.
With the examples of gust-wing interactions, how the effect of gust on {\color{black}a wing} emerges in the lift response over time is identified in an interpretable manner.
Furthermore, for the case of turbulent wake, the present model highlights structures near the wing and vortex cores as informative components based solely on the information metric without any prior knowledge of aerodynamics and length scales.
This study provides causality-based insights into a range of unsteady aerodynamic problems.
}
\end{abstract}

\vspace{-6mm}
\section*{Nomenclature}
\vspace{-5mm}


{\renewcommand\arraystretch{1.0}
\noindent\begin{longtable*}{@{}l @{\quad=\quad} l@{}}
$C_L$ & lift coefficient \\
$C_L^\prime$ & lift fluctuation \\
$c$   & chord length\\
$\cal F$   & informative mode extractor\\
$F_L$   & lift force\\
$G$   & gust ratio\\
$H$   & Shannon entropy\\
$I$   & mutual information\\
$p_{\bm q}$ & probability of the source variable\\
$p_{\bm \lambda}$ & probability of a target variable\\
$p_{{\bm \lambda},{\bm q}}$ & joint probability distribution of a target and the source variables\\
${q}$ & local flow field point\\
${q}_I$ & local informative component\\
${q}_R$ & local residual component\\
${\bm q}$ & global flow field \\
${\bm q}_I$ & global informative component \\
${\bm q}_R$ & global residual component \\
$R$ & state of the source variable\\
$R_g$ & radius of vortex gust\\
$\cal R$ & all the possible states of the source variable\\
$r$ & distance between the gust center and a point\\
$Re$ & Reynolds number\\
$S$ & state of a target variable\\
$\cal S$ & all the possible states of a target variable\\
$t$ & convective time \\
$u_i$ & velocity in the $i$th component \\
$u_\infty$ &    free stream velocity \\
$u_\theta$ &    rotational velocity \\
$w$  & spanwise velocity\\
${\bm w}$  & weights of the informative mode extractor\\
$\alpha$  & angle of attack \\
$\beta$  & weighting coefficient in cost function \\
$\omega$  & spanwise vorticity\\
$\nu$ & kinematic viscosity\\
$\rho$ & density\\
$\tau$ & wall-shear stress\\
$\theta$ & phase\\
${\bm\lambda}$ & target variable\\
${\Gamma}$ & circulation\\
${\Delta t}$ & time interval for causality estimation\\
${\Delta {\bm x}}$ & space interval for causality estimation\\


\end{longtable*}}

\section{Introduction}

\lettrine{M}{echanism} of aerodynamic-force production is of interest in studying vortical flows around air vehicles.
Aerodynamicists have analyzed the relationship between complex flow behaviors and aerodynamic forces by gazing at time traces of them measured by numerical simulations or experiments~\citep{anderson, katz2001low, leishman2006principles}.
Through this manmade process, we always aim to understand how vortical motions in a flow field contribute to the change of aerodynamic responses over time.
In other words, our eyeballs may seek what are the cause (part of a flow field of interest) and the effect (the aerodynamic response after a certain time).
This study aims to capture such a causal relationship between vortical flows and aerodynamic responses in a data-driven manner from the aspect of information.

To identify structures that primarily determine aerodynamic responses such as lift, drag, and moment, modal analysis is often considered~\citep{TBDRCMSGTU2017}.
By breaking down an unsteady flow field into a set of modes through decomposition techniques including proper orthogonal decomposition (POD)~\citep{Lumely1967}, dynamic mode decomposition~\citep{Schmid2010}, and their variants~\citep{schmidt2020guide}, the relationship between the temporal behaviors of modal coefficients or corresponding modal structures and aerodynamic forces can be examined, revealing the contribution of dominant coherent structures on aerodynamic responses.
These insights into the physics and aerodynamic systems can also be leveraged for developing reduced-order modeling, which can support the understanding and control of unsteady flows~\citep{rowley2017model,THBSDBDY2019}.
In addition to these data-driven techniques, operator-based approaches such as resolvent analysis can be considered for studying the aerodynamics by characterizing the input-output relationship across a range of frequencies and wavenumbers~\citep{MS2010,towne2018spectral}.
This mean flow-based analysis enables reasonable prediction of aerodynamic control performance based on energy amplification for a perturbation given into a flow system~\citep{yeh2019resolvent}.

Along with the above modal-analysis techniques often used for cases with a time-invariant mean flow, there exists emerging data-driven and operator-based approaches that aim to be generalized for unsteady base flows~\citep{linot2025extracting}.
For example, a dynamic network-based data-driven approach known as broadcast mode analysis provides an input-output relationship through the construction of a complex network that describes the temporal evolution of vortical interactions in unsteady flows~\citep{yeh2021network}.
This technique identifies the structures that reflect the effect of perturbation in the network constructed by the flow snapshots.
Optimally time-dependent mode decomposition has also been considered for a range of aerodynamic examples, which performs linear stability analysis for the instantaneously linearized dynamics~\citep{babaee2016minimization}.
Equipped with the time-dependent orthonormal modes derived through this linear technique, the dominant transient amplification of perturbations such as the perturbation dynamics about the time-varying mean flows can be analyzed~\citep{kern2024onset,zhong2025transient}.
A variant of resolvent analysis applicable to time-evolving flows with sparsity promotion into the modal structures has been developed for the cases with time-varying mean velocity profiles~\citep{lopez2024sparse}.
By incorporating the temporal dimension into the numerical domain via a discrete time-differentiation operator, it is possible to examine linear amplification mechanisms for time-varying base flows as a change of the space–time modal structures.
Furthermore, nonlinear machine-learning-based compression with autoencoder offers a low-rank manifold representation associated with nonlinear modes extracted from unsteady aerodynamic flow data~\citep{FT2023,fukami2024data,fukagata2025compressing}.

Aiming to capture the causal relationship between vortical structures and aerodynamic responses under either statistically stationary or time-varying mean flows, this study considers incorporating information theory into the data-driven identification of modal structures for unsteady aerodynamic flows.
This is achieved by an information-theoretic data-driven approach, informative and non-informative decomposition~\citep{arranz2024informative}.
For a given flow variable, this approach decomposes it into its informative and residual components with respect to a time-lagged target variable.
\citet{arranz2024informative} reported that it is possible to extract the informative component from a given snapshot of turbulent channel flow through data-driven optimization, which maximizes the mutual information of the informative component with the wall-shear stress at a future time stamp while minimizing the mutual information with the residual component.
With {\color{black}four} examples of separated wakes around a wing, 1. laminar periodic flows at post-stall angles of attack, {\color{black}2. numerically- and 3. experimentally-measured strong gust-wing interactions}, and 4. a turbulent flow in a spanwise-periodic domain, the present study discusses how the information-theoretic approach, which has recently been used for some problems in turbulence~\citep{lozano2022information,lopez2024linear,tanogami2024information,araki2024forgetfulness,tanogami2025scale,martinez2024decomposing}, characterizes the causality between the vortical flows and aerodynamic responses in a data-driven manner.

The present paper is organized as follows.
The information-theoretic decomposition approach is described in section~\ref{sec:methods}.
The results and discussion are presented in section~\ref{sec:res}.
Concluding remarks are offered in section~\ref{sec:conc}.

\section{Methods}
\label{sec:methods}

\begin{figure}
  \centering
    \includegraphics[width=0.65\textwidth]{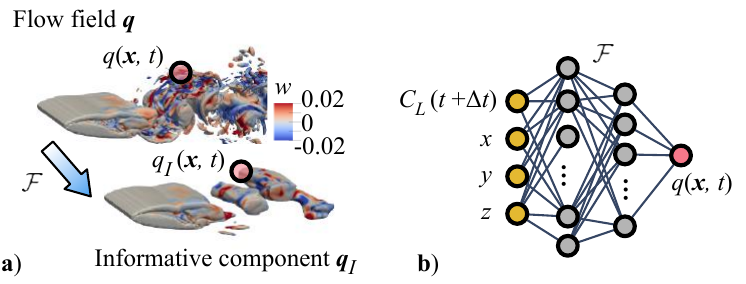}
    \caption{
    {\bf a}) An example of the given flow field ${\bm q}$ and the informative component ${\bm q}_I$.
    {\bf b}) Deep sigmoid flow.
    }
    \label{fig1}
\end{figure}

This study aims to characterize the causal relationship between vortical structures and aerodynamic responses as a modal structure.
While there is a range of conceivable measures to define the ``informative" component, this study uses the Shannon entropy for capturing the causality, following the study by~\citet{arranz2024informative}.
To explain the present approach, let us first consider extracting the informative component ${\bm q}_I({\bm x},t)$ with respect to a target variable $\bm \lambda({\bm x},t+\Delta t)$ at a future time stamp from a vortical flow ${\bm q}({\bm x},t)$ through the decomposition,
\setlength{\abovedisplayskip}{5pt}
\setlength{\belowdisplayskip}{5pt}
\begin{align}
    {\bm q}({\bm x},t) = {\bm q}_I({\bm x},t) + {\bm q}_R({\bm x},t),\label{eq:IMD}
\end{align}
where ${\bm q}_R({\bm x},t)$ describes the residual counterpart, as illustrated in Fig.~\ref{fig1}a.
While any variables can be set as a source ${\bm q}({\bm x},t)$ and a target ${\bm \lambda}({\bm x},t+\Delta t)$, this study chooses a spanwise vorticity field as ${\bm q}$ and the lift coefficient $C_L \equiv F_{L}/(0.5\rho u^2_\infty c)$ as ${\bm \lambda}$.
Here, $F_L$ is the lift force on the wing body, $u_\infty$ is the freestream velocity, $c$ is the chord length, and $\rho$ is the density.
Along with our prior knowledge of the aerodynamic relationship between lift and circulation $\Gamma$, i.e., $\Gamma \propto C_L$, the present choice of the variable combination enables us to examine how the information-theoretic approach captures important vortical structures associated with the future lift response.

To extract the informative component ${\bm q}_I$ from a given vorticity field ${\bm q}$, this study performs a modified version of the approximated informative and non-informative decomposition~\citep{arranz2024informative}.
This approach builds an informative mode extractor ${\cal F}$ based on a neural network, referred to as deep sigmoid flow~\citep{huang2018neural}, under some assumptions that relax the constraints associated with the Shannon entropy-based measures~\citep{shannon1948mathematical,arranz2024informative}, which is detailed later.
The Shannon entropy $H({\bm\lambda})$~\citep{shannon1948mathematical} for a target variable ${\bm \lambda}$ at a future time stamp is described as
\begin{align}
    H({\bm\lambda}) = -\sum_{{\bm S}\in \cal S} p_{\bm \lambda} ({\bm \lambda} = {\bm S}) \log p_{\bm \lambda} ({\bm \lambda} = {\bm S}) \geq 0,\label{eq:Ent1}
\end{align}
where $p_{\bm \lambda} ({\bm \lambda} = {\bm S})$ denotes the probability of $\bm \lambda$ in the state ${\bm S}$ and ${\cal S}$ represents all the possible states of ${\bm \lambda}$.
Now, considering the discount for the information in a given variable ${\bm q}$, the conditional Shannon information 
\setlength{\abovedisplayskip}{6pt}
\setlength{\belowdisplayskip}{6pt}
\begin{align}
    H({\bm\lambda}|{\bm q}) = -\sum_{{\bm S}\in \cal S} \sum_{{\bm R}\in \cal R} p_{{\bm \lambda},{\bm q}} ({\bm S},{\bm R}) \log \dfrac{p_{{\bm \lambda},{\bm q}} ({\bm S},{\bm R})}{p_{\bm q}({\bm R})} \geq 0\label{eq:Ent2}
\end{align}
expresses the remaining information in ${\bm \lambda}$.
Here, $p_{{\bm \lambda},{\bm q}}$ represents the joint probability distribution of ${\bm \lambda}$ and ${\bm q}$, ${\bm R}$ is a state of the source variable ${\bm q}$, and ${\cal R}$ is all the possible states of ${\bm q}$.
By taking the difference between Eqs.~\ref{eq:Ent1} and~\ref{eq:Ent2}, the amount of information shared between the given source variable ${\bm q}$ and the target variable at a future time stamp ${\bm \lambda}$ is evaluated.
This difference is called the mutual information between the variables, defined as 
\setlength{\abovedisplayskip}{3pt}
\setlength{\belowdisplayskip}{3pt}
\begin{align}
    {I}({\bm \lambda};{\bm q}) =  H({\bm\lambda}) - H({\bm\lambda}|{\bm q}) \ge 0,\label{eq:def_mut}
\end{align}
where equality is attained when ${\bm \lambda}$ and ${\bm q}$ are completely independent of each other~\citep{cover1999elements}.

We then consider decomposing the source variable ${\bm q}$ into the informative component ${\bm q}_I$ and residual counterpart ${\bm q}_R$ by Eq.~\ref{eq:IMD}.
This information-theoretic decomposition can be formulated as an optimization problem.
The condition that needs to be maximized is the mutual information between the informative component ${\bm q}_I$ and a target variable ${\bm \lambda}$ at a future time stamp ${I}({\bm \lambda};{\bm q}_I) = H({\bm \lambda})$, which is equivalent to the zero conditional Shannon entropy 
\begin{align}
    H({\bm \lambda}|{\bm q}_I) = 0,~\label{eq:DSI}
\end{align}
since ${\bm q}_I$ includes all the information of ${\bm \lambda}$, as descried in Eq.~\ref{eq:def_mut}.
Furthermore, the mutual information between the informative and residual components should be zero,
\setlength{\abovedisplayskip}{3pt}
\setlength{\belowdisplayskip}{3pt}
\begin{align}
    I({\bm q}_R;{\bm q}_I) = 0,~\label{eq:mut}
\end{align}
as they should be independent of each other.


To extract the informative component ${\bm q}_I$ by preparing a model ${\cal F}$ with the above conditions, the optimization with some assumptions in the formulation, referred to as an approximated informative and non-informative decomposition~\citep{arranz2024informative}, is performed in this study.
We particularly use two relaxations suggested in the original study~\citep{arranz2024informative}; namely,
\begin{enumerate}
    \item The source and target variables are restricted to be scalars such that the decomposition is locally performed, as illustrated in Fig.~\ref{fig1}.
    \item The minimization of Eq.~\ref{eq:mut} is sought by incorporating it into the cost function rather than strictly constraining it.
\end{enumerate}
The former alleviates the issue of the curse of dimensionality from flow field data ${\bm q}$ in computing joint probability distributions.
This becomes especially challenging in considering flow fields as they often possess a high degree of freedom.
Hence, this first relaxation to employ the local variable $q$ instead of the global one ${\bm q}$ offers the tractability of the approach even for flow field data, although such local-based approaches often encounter difficulties in producing spatially continuous flow fields~\citep{FNKF2019}.
The latter reduces the complexity of optimization as it is generally challenging to find the unknown probability distribution arising from the residual component ${\bm q}_R$.
Equipped with these conditions, the approximated informative and non-informative decomposition is formulated as an optimization for the weights $\bm w$ in an informative mode extractor ${\cal F}$ to produce the best, smooth output of ${q_I}$,
\begin{align}
    {\bm w}^* = {\rm argmin}_{\bm w} ||q-q_I||_2 + \beta ||I(q_R;q_I)||_2,
\end{align}
where $\beta$ balances the reconstruction loss and the mutual information loss.
The resulting informative component is described as
\vspace{-3mm}
\begin{align}
    {q_I} = {\cal F}(\lambda; {\bm w}),
\end{align}
with an assumption that the mapping ${\cal F}$ between the variables is invertible.
The weighting coefficient $\beta$ in this study is determined based on the L-curve analysis~\citep{hansen1993use}, which finds an appropriate regularization parameter of the cost function.
We will also perform a parametric study of $\beta$ with an example of vortex-gust airfoil interaction to observe how the resulting modal structures vary depending on $\beta$.

The original study~\citep{arranz2024informative} considered a velocity $u_i({\bm x},t)$ at a point ${\bm x}$ as ${q}$ and a wall-shear stress $\tau({\bm x}-\Delta {\bm x},t+\Delta t)$ as $\lambda$, where $\Delta {\bm x}$ and $\Delta t$ are respectively space and time intervals, for a turbulent channel flow.
While both variables are a function of space in their case, our target variable of the lift coefficient $C_L(t+\Delta t)$ is spatially global.
Hence, an informative mode extractor ${\cal F}$ in this study needs to additionally take variables with the information of spatial coordinate ${\bm x}$, enabling the model ${\cal F}$ to recognize where the location of interest is in decomposing the given variable ${q({\bm x},t)}$.
In response, the present mode extractor ${\cal F}$ receives the spatial coordinate information of ${\bm x}$ in addition to a target variable ${\lambda} = C_L(t)$, as illustrated in Fig.~\ref{fig1}b.
A modified version of approximated informative and non-informative mode decomposition for the present aerodynamic problems is hence expressed as 
\begin{align}
    q_I = {\cal F}(C_L(t+\Delta t),{\bm x}),
    \label{eq:output}
\end{align}
with the optimization,
\begin{align}
    {\bm w}^* &= {\rm argmin}_{\bm w}||q({\bm x},t)-q_I({\bm x},t)||_2 + \beta ||I({q}_R({\bm x},t);{q}_I({\bm x},t))||_2 \nonumber\\
    & = {\rm argmin}_{\bm w}||q({\bm x},t)-{\cal F}(C_L(t+\Delta t),{\bm x})||_2 + \beta ||I(q({\bm x},t)-{\cal F}(C_L(t+\Delta t),{\bm x});{\cal F}(C_L(t+\Delta t),{\bm x})||_2,\label{eq:cost}
\end{align}
which learns the causality from the informative component ${\bm q}_I({\bm x},t)$ to the lift coefficient $C_L({\bm x},t+\Delta t)$ over the time interval $\Delta t$.
In other words, the modal structures ${\bm q}_I$ are expected to vary with the choice of $\Delta t$ since the extent to which the given vortical structures are evaluated as "informative" depends on the time window between the variables ${q}_I(t)$ and $C_L(t+\Delta t)$.
Furthermore, the current approach produces time-varying modes as the mode is given for each snapshot of ${\bm q}(t)$.

As illustrated in Fig.~\ref{fig1}b and described in Eq.~\ref{eq:output}, the current model ${\cal F}$ takes the local coordinates $x$ and the lift response $C_L$ at a future time stamp ${t+\Delta t}$ and provides the informative component $q_I$ at the current time $t$.
To construct such an informative mode extractor ${\cal F}$, we use a fully-connected neural network (multi-layer perceptron)~\citep{RHW1986} with some constraints, refereed to as the deep sigmoidal flow~\citep{huang2018neural}.
This multi-layer perceptron is composed of bijective activation functions with positive weights.
With these constraints, the deep sigmoidal flow achieves a bijective transformation, enabling the solution of ${q_I}$ produced by the mode extractor ${\cal F}$ to satisfy the constraint of the discounted Shannon information $H(C_L({\bm x,t+\Delta t}),q_I({\bm x},t)) = 0$ in Eq.~\ref{eq:DSI}.
The mutual information in this study is computed with the Gaussian kernel density estimator.
The optimization in Eq.~\ref{eq:cost} is performed with the Adam algorithm~\cite{Kingma2014}.
Further details on the properties of deep sigmoidal flow with respect to the informative mode decomposition are referred to \citet{arranz2024informative}.

\vspace{-5mm}
\section{Results}
\label{sec:res}

In this section, we apply a modified version of the approximated informative and non-informative mode decomposition, hereafter called informative mode decomposition or IMD for brevity, to {\color{black}four} examples of separated flows, covering a range of spatiotemporal aerodynamic characteristics.
The first example of laminar airfoil flow across the angles of attack exhibits periodic wake shedding while its frequency varies depending on the angle.
The second case of {\color{black}numerically-simulated} vortex-gust airfoil interactions presents highly unsteady transient dynamics with a large excitation of aerodynamic forces during short time.
{\color{black}While the third case of experimental data for strongly-disturbed flow around a wing also exhibits transient dynamics, these data include measurement noise.}
Furthermore, the {\color{black}fourth} problem of a turbulent separated wake involves the three-dimensionality of vortical structures with quasi-periodicity over time.
Along with the current problem settings, let us examine how the present data-driven approach extracts physics for unsteady aerodynamic problems from an informatics point of view.


\subsection{Example 1: periodic laminar airfoil wake shedding at post-stall angles of attack}
\label{sec:ex1}

We first consider a flow around a NACA0012 airfoil at a chord-based Reynolds number $Re = u_\infty c/\nu = 100$, where $\nu$ is the kinematic viscosity, with angles of attack $\alpha\in [30^\circ, 50^\circ]$.
Direct numerical simulations are performed to produce the present data sets~\citep{cliff1,cliff2,FT2023}.
The computational domain is set over $(x, y)/c \in [-15, 30] \times [-20, 20]$ with the leading edge of the wing positioned at the origin. 
The present simulations have been carefully verified and validated with previous studies~\citep{ZFAT2023,kurtulus2015unsteady,liu2012numerical,di2018fluid}.
Further details on the simulation setup are referred to as Fukami and Taira~\cite{FT2023}.

\begin{figure}[t]
  \centering
    \includegraphics[width=0.75\textwidth]{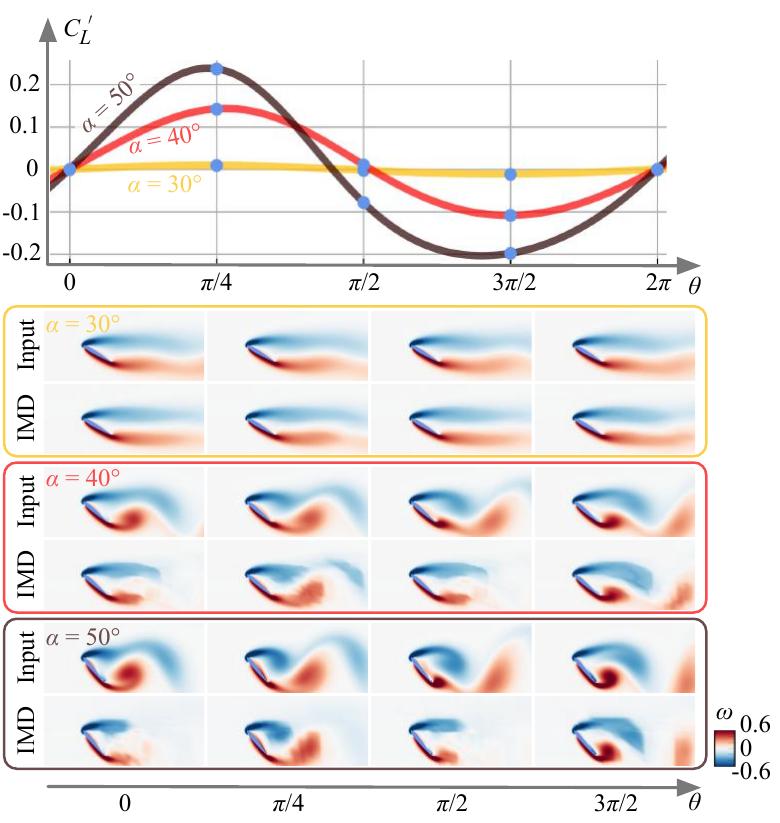}
    \caption{
    Informative mode decomposition of laminar airfoil wake.
    Vorticity snapshots and decomposed fields over phase across the angle of attack are shown with the lift fluctuation.
    }
    \label{fig2}
\end{figure}

At the current Reynolds number with $\alpha \gtrsim 30^\circ$, wakes present periodic shedding, producing a limit-cycle oscillation in the lift response, as shown in Fig.~\ref{fig2}.
By increasing the angle of attack, the vortex roll-up near the leading and trailing edges is more clearly observed, exhibiting larger vortical structures with higher shedding frequency.
Correspondingly, the lift fluctuation $C_L^\prime$ from their mean values becomes larger as the angle of attack increases.
Hence, this example enables a comprehensive analysis with respect to the size of vortical structures and the level of lift fluctuation over time.
For the current information-based analysis, the parameters of the convective time window $\Delta t$ and the weighting coefficient $\beta$ are set to 0.0085 and 0.01, respectively. 
Note that the dependence of modal structures on the choice of these parameters is examined with the second example of vortex-airfoil interactions later.

The present informative mode decomposition is applied to the periodic shedding examples, as shown in Fig.~\ref{fig2}.
Here, vorticity snapshots ${\bm q}$ and resulting decomposed fields ${\bm q}_{I}$ are presented over phase $\theta = \tan^{-1}(\dot{C_L}(t)/C_L(t))$ of $0, \pi/2, \pi$, and $3\pi/2$ of the current periodic dynamics for lift~\citep{kawamura2022adjoint}.
Although the informative component of the flow field with $\alpha = 30^\circ$ simply aligns with the vortical structures over time when the lift fluctuation is small, the behavior of the informative modes varies over time by increasing the angle of attack.
For $\alpha \geq 40^\circ$, while the structures near the leading and trailing edges are only responsible at $\theta = 0$ and $\pi/2$ when the lift fluctuation is small, the vortex core in downstream becomes highlighted at $\theta = \pi/4$ and $3\pi/2$ with a large lift fluctuation.
This indicates that in evaluating whether vortical structures are informative or not, the current model based on the information successfully captures the relationship of $\Gamma \propto C_L$ rather than solely outputting large vortex cores.
In addition, the informative modal structures are more localized and their interests shift toward the region near the wing as $\alpha$ further increases to $50^\circ$.
This coincides with observations made in previous studies that the region near the wing becomes highly responsible at high $\alpha$ for modifying the phase dynamics of lift~\citep{kawamura2022adjoint,godavarthi2023optimal,iima2024optimal}. 
Such aerodynamically interpretable modal structures are produced in a time-varying manner.

\subsection{Example 2: strong vortex-gust airfoil interactions}
\label{sec:ex2}

To examine the current approach for transient aerodynamic problems, let us then consider a strong vortex gust-airfoil interaction.
This type of aerodynamic scenario occurs for small-sized aircraft flying in a range of environments such as urban canyons, mountainous areas, and turbulent wakes created by ships.
As the influence from gusts becomes non-negligible for such small-scale air vehicles due to their relative size, characterizing aerodynamics under such strong gust encounters is critical toward achieving stable flights under severe atmospheric conditions~\citep{jones2022physics,jones2021overview,FT2023}.

By taking the baseline (undisturbed) flow at $\alpha = 40^\circ$ in the first example, a strong vortex gust is introduced upstream of a NACA0012 airfoil at $x/c = -2$ and $y/c = 0.1$.
The disturbance is modeled using a Taylor vortex~\citep{taylor1918dissipation} with a rotational velocity profile of
\setlength{\abovedisplayskip}{10pt}
\setlength{\belowdisplayskip}{10pt}
\vspace{-4mm}
\begin{align}
    u_{\theta}=u_{\theta, \rm{max}}{\dfrac{r}{R_g}}{\rm exp}\left[\dfrac{1}{2}\left({1-\dfrac{r^2}{R_g^2}}\right)\right] ,
\end{align}
where $R_g$ is the radius at which $u_{\theta}$ reaches its maximum velocity $u_{\theta, \rm{max}}$.
This study considers cases with gust ratio $G \equiv u_{\theta,\text{max}}/u_\infty \in [-2,2]$ and gust diameter $D \equiv 2R_g/c = 0.5$.

Since the case of $\alpha = 40^\circ$ in the first example is taken as the baseline undisturbed condition, we use the same computational domain for the current simulation as that used for the first periodic shedding cases.
The simulations with vortex gust encounters have also been verified and validated with previous studies~\citep{ZFAT2023,FT2023,fukami2024data}.
Further details on the simulation setup are referred to as Fukami and Taira~\cite{FT2023}.
The baseline parameters of the convective time window $\Delta t$ and the weighting coefficient $\beta$ in the second example are set to 0.0085 and 0.01, respectively, although the dependence of the dynamic modal behavior on the parameter choice will be examined later.

The gust ratio $G$, characterizing the relative influence of aircraft on disturbance, is particularly important for studying vortex-gust airfoil interactions.
Small-sized air vehicles encounter flight conditions of $|G| > 1$ under the above-mentioned violent airspace, which is traditionally avoided~\citep{jones2022physics}.
Furthermore, due to strong vortex impingement, large and transient excitation of aerodynamic forces occurs within a very short duration of 1-2 convective time~\citep{FT2023}.
Hence, this second example with the current gust ratio of $|G|\geq 1$ is regarded as highly unsteady transient conditions.

\begin{figure}[t]
  \centering
    \includegraphics[width=0.8\textwidth]{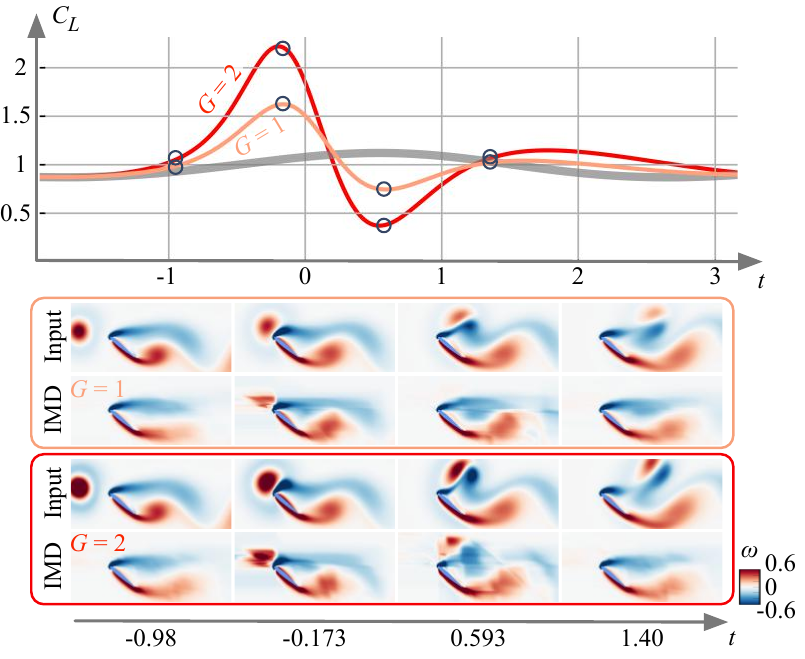}
    \caption{
    Informative mode decomposition of strong vortex-airfoil interactions for positive gust.
    Vorticity and decomposed fields are shown with the lift coefficient (gray line: the undisturbed case).
    }
    \label{fig3}
\end{figure}


Here, we apply the present informative mode decomposition to strong vortex-airfoil interactions with a positive gust, as presented in Fig.~\ref{fig3}.
The convective time for the vortex-airfoil interaction case is set to zero when the center of the vortex reaches the leading edge of the wing.
For the cases with positive gusts, lift first increases from the undisturbed response (gray line).
After the vortex impingement, the interaction between the positive gust and the structure possessing a negative vorticity near the leading edge of the airfoil causes massive separation, reducing the lift over $0 < t < 1$. 
The fluctuation level from the undisturbed state becomes large as the gust ratio increases.


We have found that the present information-based analysis captures how the effect of the gust on the aerodynamic response varies in a time-varying manner.
When the lift starts to deviate from the undisturbed response at $t = -0.98$, the gust does not appear in the informative mode as its influence on the lift is weak. 
This suggests that the lift at this moment is primarily determined by the pre-existing structures around a wing rather than the vortex gust.
The gust is then assessed as ``informative" near the first lift peak at $t=-0.173$.

Once the negative vorticity near the leading edge is pinched off by a gust with $G=1$ at $t=0.593$, rather than by the gust itself, the modified structures around a wing by the gust are assessed as informative to determine the lift.
Although the gust-like structure emerges in the mode by increasing $G$ to 2, its magnitude is under-evaluated compared to the input data. 
These results indicate that the structures deformed due to the vortex gust are important for lift dynamics from the aspect of information.

\begin{figure}[t]
  \centering
    \includegraphics[width=0.8\textwidth]{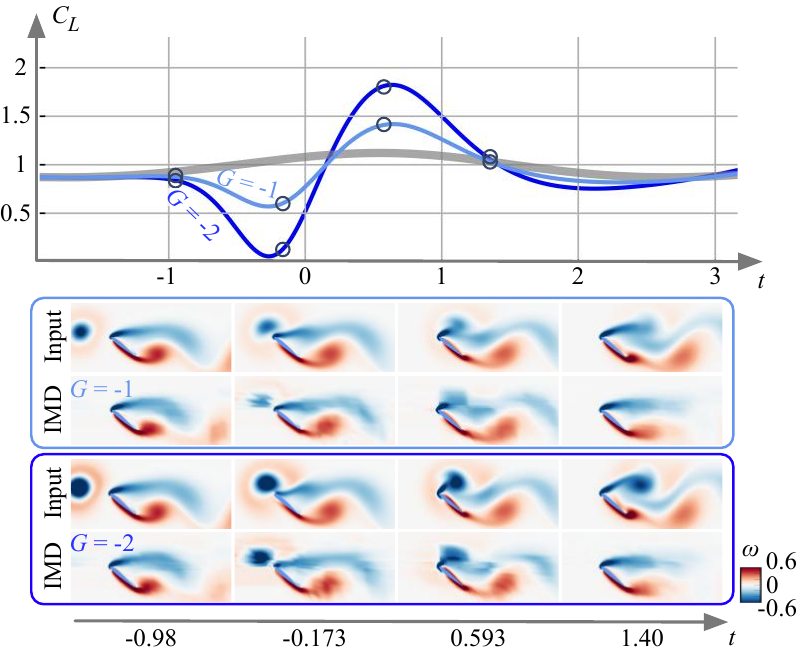}
    \caption{
    Informative mode decomposition of strong vortex-airfoil interactions for negative gust.
    Vorticity and decomposed fields are shown with the lift coefficient (gray line: the undisturbed case).
    }
    \label{fig4}
\end{figure}

We also consider the present informative mode decomposition for strong vortex-airfoil interactions with a negative gust, as shown in Fig.~\ref{fig4}.
In contrast to the cases with a positive gust, negative vortex disturbances first decrease the lift, which is subsequently recovered towards that of the undisturbed baseline response.
The vortex gust and the pre-existing structure near the leading edge with the same sign of negative vorticity transiently interact and merge, causing large deformation of vortical structures in a wake.
As well as the positive gust impingement, the fluctuation from the undisturbed lift increases as $|G|$ becomes large.

Similar to the positive gust cases, the gust originally does not show up in the mode fields at $t=-0.98$ while starting to emerge over time, corresponding to its influence on lift.
At $t=0.593$ when the gust interacts with the structure around the leading edge, the mode for $G=-1$ assesses the entire deformed structure with negative vorticity as informative while that for $G=1$ in Fig.~\ref{fig3} does not include a gust in the mode.
This difference between the positive and negative vortex-gust interactions also suggests that the modification of structures around the leading edge due to the vortex impingement primally affects to the sharp fluctuation of the lift response rather than the gust itself.

\begin{figure}[t]
  \centering
    \includegraphics[width=0.8\textwidth]{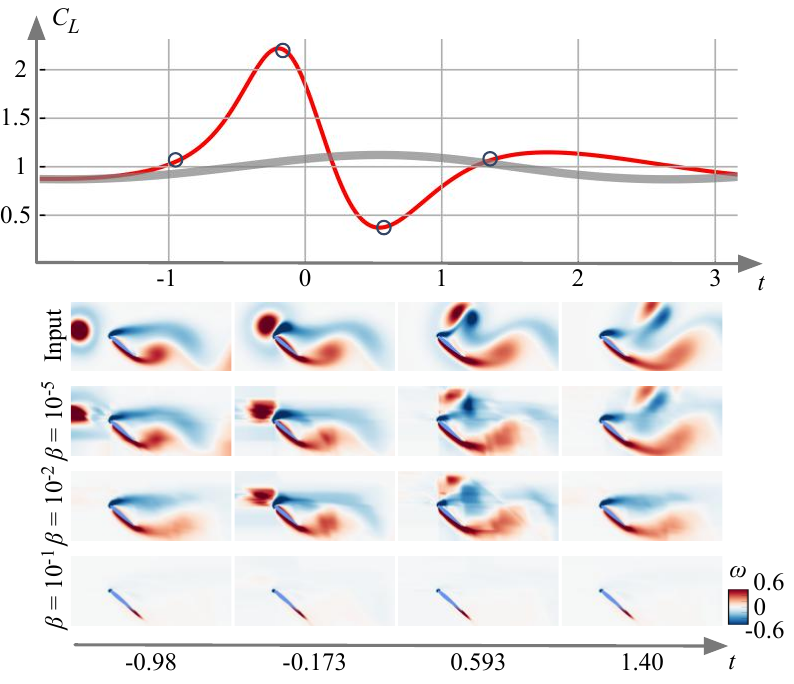}
    \caption{
    Dependence of informative modal structures on the weighting parameter.
    Vorticity and decomposed fields are shown with the lift coefficient (gray line: the undisturbed case).
    }
    \label{fig5}
\end{figure}

The present extraction of the informative component from a given flow snapshot is achieved by introducing the cost based on mutual information in Eq.~\ref{eq:cost}.
While the optimal choice of $\beta$ can be sought with several techniques including the L-curve analysis~\cite{hansen1993use} used in this study, how the mode identification would vary by changing the weight $\beta$ may also be of interest.
Here, let us examine the dependence of informative modal structures on the weighting parameter $\beta$ of the cost function. 
A strong vortex-airfoil interaction with $G=2$ is considered as an example case, as presented in Fig.~\ref{fig5}.
With a small $\beta$ of $10^{-5}$, the modes produced by the model are almost identical since the model performs a sole regression.
As $\beta$ increases, the modal structures become different from the input data, extracting the informative components with $\beta = 0.01$.
Once the penalization becomes too large such as the case with $\beta = 0.1$, the first term of the cost function is no longer reduced, thereby producing excessively localized structures at the leading and trailing edges, as shown in Fig.~\ref{fig5}.
In other words, physically meaningful modal structures can be extracted using the current technique with an appropriate choice of the weighting parameter.

\begin{figure}[t]
  \centering
    \includegraphics[width=0.8\textwidth]{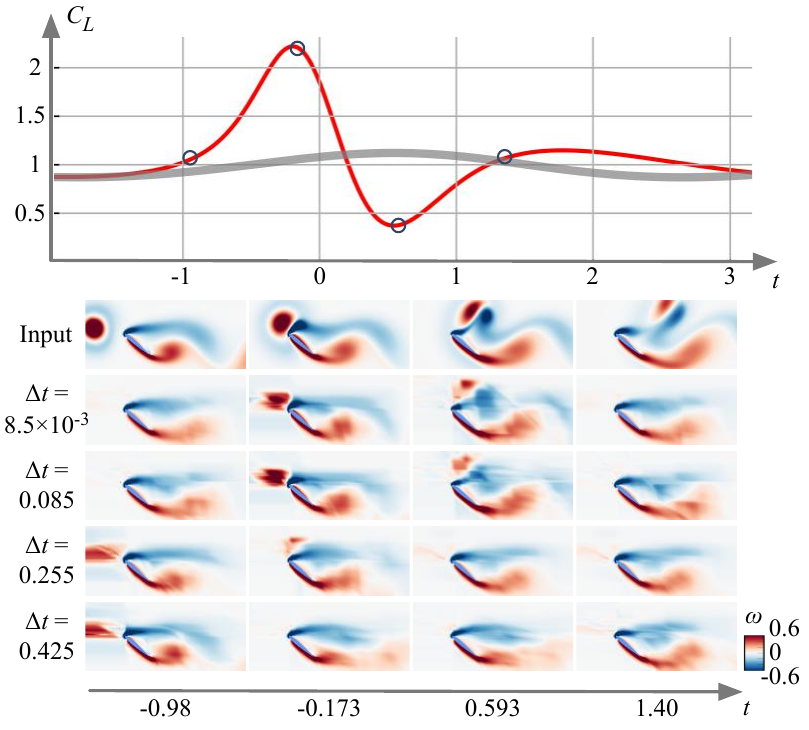}
    \caption{
    Dependence of informative modal structures on the time window.
    Vorticity and decomposed fields are shown with the lift coefficient (gray line: the undisturbed case).}
    \label{fig6}
\end{figure}

To further analyze whether the present formulation captures the causal relationship between vortical structures and aerodynamic responses, the dependence of informative modal structures on the convective time window $\Delta t$ is examined, as shown in Fig.~\ref{fig6}.
We again consider a strong vortex-airfoil interaction with $G=2$ for this analysis.
Since the vortical structures and lift responses in this case clearly vary over time in a transient manner, this example is a right testbed to observe the effect of time-window setting.

With the baseline time window of $\Delta t = 0.0085$, the gust does not appear in the mode since the lift right after the short time duration is primarily determined by the structures around a wing rather than a gust, as discussed above.
As $\Delta t$ increases, the delay effect appears in the time-varying modes.
Gust-like structures emerge with a larger time window of $\Delta t = 0.255$ and 0.425 at $t = -0.98$ since the gust influences when the lift response deviates from the undisturbed case.
In contrast, the gust is no longer seen at $t = -0.173$ and 0.593 with a larger time window due to the convection process.
These observations suggest that the model learns causality over the convection process of strong vortex-airfoil interactions and successfully extracts it as a time-varying mode.

\subsection{{\color{black}Example 3: experimental measurements of large-amplitude transverse gust encounter}}
\label{sec:ex3}

\begin{figure}
  \centering
    \includegraphics[width=0.75\textwidth]{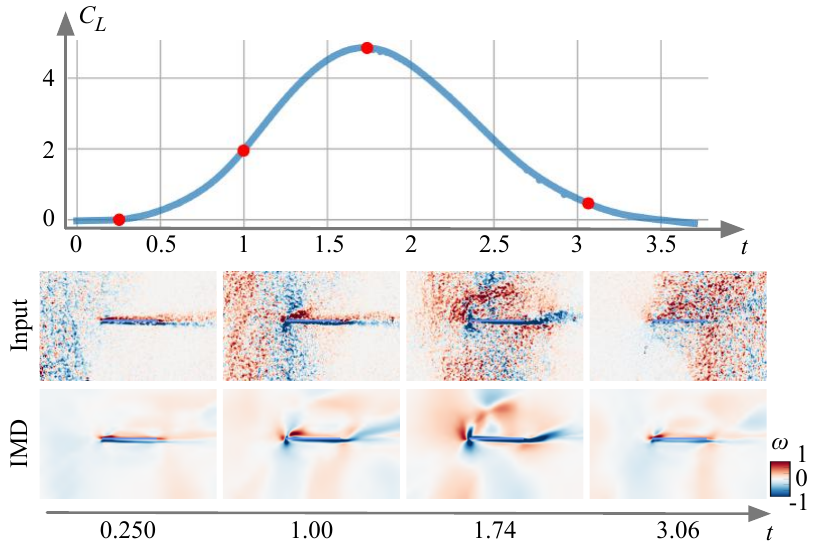}
    \caption{
    {\color{black}
    Informative mode decomposition of experimentally-measured large-amplitude transverse gust encounter at $Re = 20,000$.
    Vorticity fields and resulting decomposed fields are shown with the lift coefficient.}
    }
    \label{fig_Exp}
\end{figure}

{\color{black}
To examine the robustness of the present approach against experimental flow measurements at a high Reynolds number, this section considers a flow around a flat-plate wing at a constant towing speed and angle of attack $\alpha$ passing through a large-amplitude transverse gust at a gust ratio $G = 1.5$.
The data set is made available by \citet{towne2023database}.
Experiments were performed in the University of Maryland free-surface water towing tank, and a flat-plate wing made of glass was used in measuring flow fields by particle image velocimetry.
The time trace of aerodynamic forces is also collected along with that of flow fields, as shown in Fig.~\ref{fig_Exp}.
The flat plate here is set to a fixed $\alpha = 0^\circ$ and encounters a transverse gust with a sine-squared profile. 
The chord-length-based Reynolds number is 20,000.
Both the vorticity field and the lift coefficient used in the present study are ensemble-averaged over eight and five individual runs, respectively. 
Further details on the experimental setups are referred to previous studies~\citep{biler2021experimental,andreu2020effect,towne2023database}.
With the present strong gust ratio $G\geq 1$, this problem setup enables us to analyze the applicability of the approach for highly unsteady transient dynamic data measured under gusty and noisy conditions.

Let us perform the present informative mode decomposition for the time series of experimental measurements, as exhibited in Fig.~\ref{fig_Exp}.
Here, the parameters of the convective time window $\Delta t$ and the weighting coefficient $\beta$ are set to 0.005 and 0.01, respectively. 
With the approach of the transverse gust, the lift coefficient sharply increases within sole two convective time and eventually is relaxed to the statistically steady state.
Corresponding to this fast and large excitation of aerodynamic forces, the large-scale flow separation occurs due to the formation of a leading-edge
vortex, as seen for $1\leq t \leq 1.74$.
They eventually become turbulent fine structures and convect over a wing.

Similar to the numerical example of vortex-airfoil interactions, the gust does not appear in the informative modal field at the early stage of $t=0.250$ as the lift at that moment is primarily determined by the flow around a wing.
Once the transverse gust begins to interact with a wing, the separated structures near the leading edge are extracted as the informative components.
At $t=1.74$ presenting the lift peak, largely separated structures on the suction side and the newly-generated structure possessing negative vorticity near the trailing edge are highlighted in the informative modal field.
Once the flow becomes turbulent and starts to dissipate at $t=3.06$, fine-scale structures with small contribution to the lift response are discarded from the informative mode.
These observations indicate that structures deformed due to the gust-wing interaction mainly affect the lift dynamics, which coincides with our findings through the numerical example of strong vortex-airfoil interactions.
This type of high Reynolds number aerodynamic flow can also be analyzed with the present informative mode decomposition.

Furthermore, noteworthy here is that the current approach offers time-varying informative modes while performing denoising.
Such noisy components in experimental measurements often hinder the extraction of dominant structures in a data-driven manner, necessitating special care in the formulation such as sparsity promoting~\cite{scherl2020robust}.
The successful extraction of clean modal structures indicates that the present information-theoretic framework finds the causal relationship between vortical structures and aerodynamic responses while removing noisy components that have no causal effects on lift dynamics, exhibiting the applicability to a range of aerodynamic flow scenarios.  
}

\subsection{Example 4: turbulent separated wake in a spanwise-periodic domain}
\label{sec:ex4}

\begin{figure}[t]
  \centering
    \includegraphics[width=0.85\textwidth]{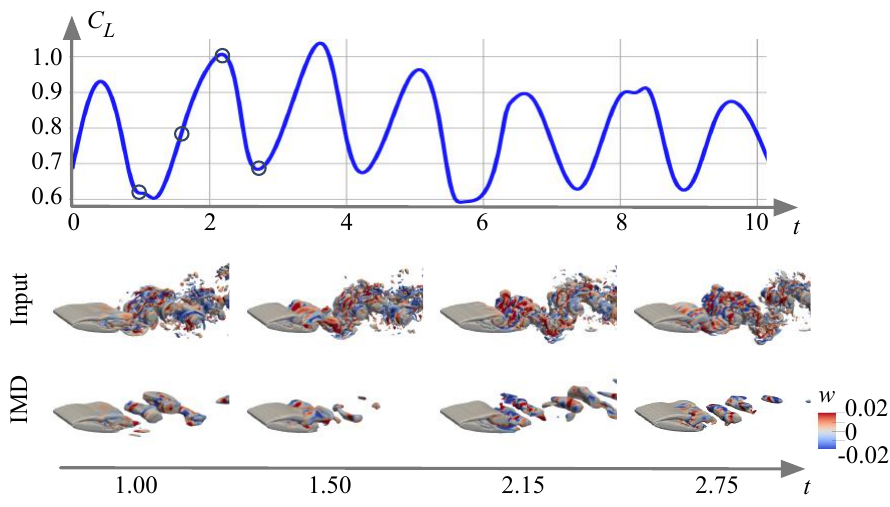}
    \caption{
    Informative mode decomposition of turbulent separated wake at $Re = 5,000$.
    Vorticity isocontours and decomposed fields colored by spanwise velocity are shown with the lift coefficient.
    }
    \label{fig7}
\end{figure}


To examine the applicability of the present approach for turbulent aerodynamic flows with the three-dimensionality of vortical structures, let us also consider a turbulent separated wake at $Re = 5,000$.
A large-eddy simulation (LES) of a NACA0012 airfoil at the angle of attack $\alpha=14^\circ$ is performed to prepare the present data set.
An incompressible flow solver {\it Cliff} (in {\it CharLES} software package) based on the finite-volume formulation with second-order accuracy in time and space~\cite{cliff1,cliff2} is used for the present simulation.
The computational domain is set over $(x,y,z)/c\in[-20,25]\times[-20, 20] \times [0,1]$ with the leading edge of the wing positioned at the origin.
With spanwise extent of $1c$, we consider a spanwise periodic domain for the present simulation, capturing the three-dimensional flow structures~\cite{rolandi2024invitation,rolandi2025biglobal,FST2025}.
To provide turbulent closure in the present LES, we use the Vreman subgrid-scale model~\cite{vreman2004eddy}.
The current simulation has been verified and validated with previous studies~\cite{rolandi2024invitation,rolandi2025biglobal,FST2025}.
Further details on the simulation setup are given by Fukami et al.~\cite{FST2025}.
The parameters of the convective time window $\Delta t$ and the weighting coefficient $\beta$ are set to 0.05 and 0.01, respectively.

The present informative mode decomposition is applied to the turbulent wake example, as shown in Fig.~\ref{fig7}.
The present turbulent separated wake possesses the three-dimensionality of vortical structures while exhibiting quasi-periodicity over time, as seen from the lift response.
A similar behavior of the localization of modal structures depending on the lift fluctuation to the laminar wake example is seen, as observed by comparing the modes at $t=1.50$ and others.
Noteworthy here is that the current information-based technique extracts the dominant vortex cores as ``informative" while discarding fine structures that contribute less on lift.
While this is somewhat similar to outcomes from spatial filtering approaches~\cite{fujino2023hierarchy,FST2025}, we note that the current approach only relies on flow data and information metrics without being explicitly given length-scale information.
This suggests that what is traditionally known as the relationship between a global quantity of lift and large-scale vortical motion in aerodynamics may further be evident from the informatics point of view.

\begin{figure}[t]
  \centering
    \includegraphics[width=\textwidth]{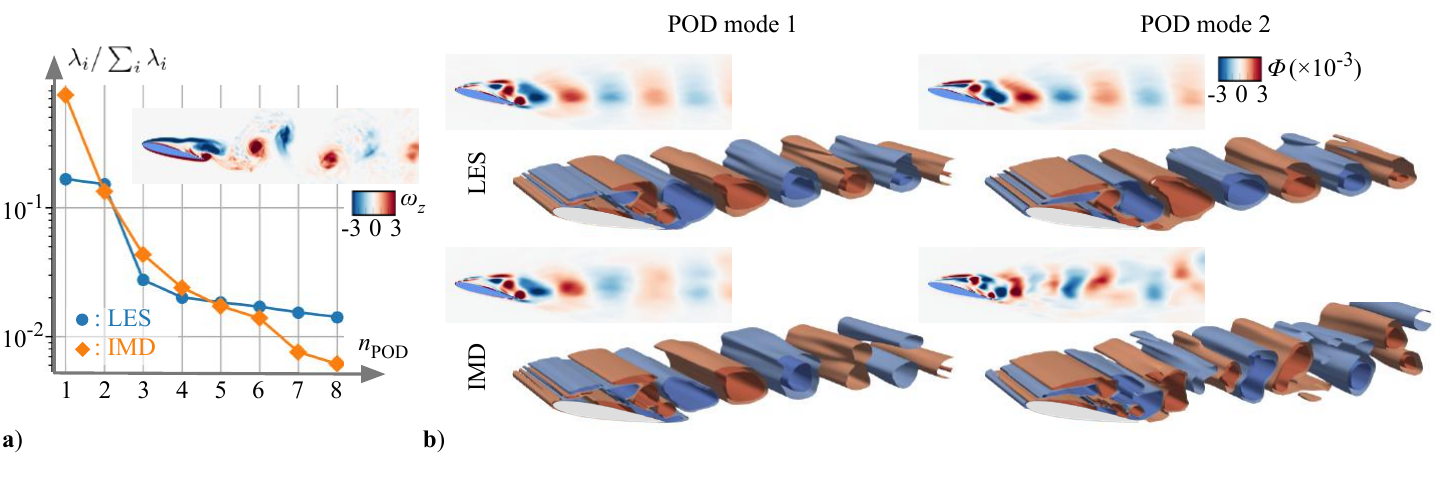}
    \caption{
    POD for LES data and informative modes of the turbulent wake.
    {\bf a}) Normalized kinetic energy.
    {\bf b}) The first and second dominant POD modes.
    }
    \label{fig8}
\end{figure}

To highlight the difference between a regular, linear decomposition and the present nonlinear information-based technique, we lastly perform POD for a time series of the original LES fields and the present decomposed fields by IMD, as summarized in Fig.~\ref{fig8}.
This analysis enables the identification of dominant vortical patterns residing in both the LES and the informative fields over time.
POD modes 1 and 2 from the LES fields are paired in the plot of the normalized kinetic energy, capturing the dominant vortex shedding.
In contrast, there is no longer a paired mode for the informative counterpart, as observed from the second POD modes and the energy plot, although the first mode from the informative components presents a similar structure to that from LES.
This is likely because the appearance of the large vortex cores downstream for the informative modes depends on the fluctuation of lift over time while they always exist in the original LES field.

Although both POD and the information-based technique primarily capture large-scale vortical structures, how each decomposition method extracts these structures over time is different due to the distinct norms employed.
POD identifies dominant structures based on their energetic contribution, typically through the minimization of $L_2$ error or the maximization of kinetic energy.
In contrast, the present data-driven, information-based technique extracts structures based on their relevance to the future lift response --- that is, how informative they are in relation to it over a certain time interval.
As a result, the structures regarded as dominant evolve differently over time, leading to notable differences in the outcomes of POD analysis.
These findings suggest the potential of the current nonlinear information-theoretic framework to reveal the causal relationship between various variables for a range of unsteady aerodynamic problems.

\section{Conclusions}
\label{sec:conc}

This study performed an information-theoretic mode decomposition for separated wakes around a wing using a data-driven framework based on deep sigmoidal flow.
The present approach extracts causally informative flow structures from an instantaneous snapshot with respect to their influence on future aerodynamic responses, enabling a data-driven time-varying modal analysis.
Through {\color{black}four} representative cases around {\color{black}a wing --- periodic laminar wake, numerically- and experimentally-measured strong gust-wing interactions}, and turbulent wake --- we showed that the method reveals informative vortical structures associated with the time-varying lift force. 
The current results highlight the potential of the present method to analyze causality-driven modal structures in complex unsteady aerodynamic flows.

The current mesh-free technique of deep sigmoidal flow was performed locally at each point in a given snapshot.
It is not only adaptable but also computationally tractable for a range of aerodynamic scenarios. 
However, this local approach introduces some level of spatial discontinuities in the resulting modal structures, as also observed in the present study.
Although performing mutual information-based optimization for the entire global field possesses challenges due to a large degree of freedom (the curse of dimensionality), the extension of the present formulation for convolutional neural network-based models may be promising for obtaining spatially continuous profiles in the mode while accounting for geometric flow features in the learning process.
Furthermore, the theoretical connection between the present information-based technique and other data-driven approaches, such as Shapley additive explanations (SHAP)~\cite{cremades2024identifying,cremades2025additive}, would be of interest to further deepen the understanding of data-driven causality analysis for unsteady flows.
{\color{black}As recently performed with the SHAP technique, one can consider performing reinforcement learning-based flow control based on the informative mode as a reward~\cite{beneitez2025improving}.
Examining the learned representation in the network may also help further assess what is learned as the causality between vortical structures and aerodynamic responses in an interpretable manner.}
Equipped with these enhancements, the current nonlinear, information-theoretic framework may serve as a foundation for extracting causality-based insights into a range of unsteady aerodynamic problems.

\section*{Funding Sources}
RA acknowledges the support from JSPS KAKENHI Grant Number JP24K22942.


\bibliography{references.bib}

\end{document}